\documentclass[prl,aps]{revtex4}
\usepackage{graphicx}

\begin{document}

\title{The antiferromagnetic Ising chain \\
in a mixed transverse and longitudinal magnetic field}

\author{A.A.Ovchinnikov}
\author{D.V.Dmitriev}
 \email{dmitriev@deom.chph.ras.ru}
\author{V.Ya.Krivnov}
\affiliation{Joint Institute of Chemical Physics of RAS, Kosygin
str.4, 117977, Moscow, Russia.}

\author{V.O.Cheranovskii}
\affiliation{Institute of Chemistry, Kharkov National University,
61077, Kharkov, Ukraine}

\date{\today}

\begin{abstract}
We have studied the antiferromagnetic Ising chain in a transverse
magnetic field $h_{x}$ and uniform longitudinal field $h_{z}$.
Using the density matrix renormalization group calculation
combined with a finite-size scaling the ground state phase diagram
in ($h_{x},h_{z}$) plane is determined. It is shown that there is
an order-disordered transition line in this plane and the critical
properties belong to the universality class of the two-dimensional
Ising model. Based on the perturbation theory in $h_{z}$ the
scaling behavior of the mass gap in the vicinity of the critical
point ($h_{x}=1/2,h_{z}=0$) is established. It is found that the
form of the transition line near the classical multicritical point
($h_{x}=0,h_{z}=1$) is linear. The connection of the considered
quantum model with the quasi-one-dimensional classical Ising model
in the magnetic field is discussed.
\end{abstract}

\maketitle

\section{Introduction}

Recently, the study of the field-induced effects in
low-dimensional quantum spin systems has been attracting much
interest from theoretical and experimental points of view
\cite{YbAs},\cite{Affleck},\cite{kufo},\cite{Kohgi}. For the
system with an anisotropy of exchange interactions the magnetic
properties essentially depend on the direction of the applied
magnetic field. For example, the behavior of the one-dimensional
antiferromagnetic $XXZ$ model in a transverse magnetic field is
drastically different in comparison with the case of the
longitudinal field. In particular, the transverse field induces
the staggered magnetization in the perpendicular direction and the
continuous phase transition takes places at some critical field
\cite{mori},\cite{hieda},\cite{we},\cite{dutta}. This effect has
been observed in quasi-one-dimensional antiferromagnet ${\rm
Cs}_{2}{\rm CoCl}_{4} $ \cite{kenz}, where the magnetic field has
both the transverse and longitudinal components. Therefore, it is
important to study the properties of the antiferromagnetic
$s=\frac{1}{2}$ $XXZ$ model in mixed transverse and longitudinal
magnetic fields
\begin{equation}
H =\sum (S_{n}^{x}S_{n+1}^{x}+S_{n}^{y}S_{n+1}^{y}+\Delta
S_{n}^{z}S_{n+1}^{z}) -h_{x}\sum S_{n}^{x}-h_{z}\sum S_{n}^{z}
\label{HXXZ}
\end{equation}

We consider the most simple case of this model -- the
antiferromagnetic Ising chain given by the Hamiltonian
\begin{equation}
H=\sum S_{n}^{z}S_{n+1}^{z}-h_{x}\sum S_{n}^{x}-h_{z}\sum S_{n}^{z}
\label{H}
\end{equation}

In spite of the simple form of this Hamiltonian it cannot be
solved exactly. In \cite{sen} the ground state phase diagram of
this model has been investigated using numerical diagonalization
of finite systems and finite size scaling procedure. It was found
\cite{sen} that the transition line between the ordered and
disordered phases exists and it was assumed that the model belongs
to the universality class of the two-dimensional Ising model.

We note that the ferromagnetic Ising chain in the mixed fields has
been studied intensively \cite{fog},\cite{zam},\cite{fateev}.
Though the ferromagnetic and the antiferromagnetic Ising chains in
the transverse magnetic field are equivalent, properties of these
two models are very different at $h_{z}\neq 0$. (In fact, the
model (\ref{H}) can be transformed to the ferromagnetic chain but
in a staggered longitudinal field.) For example, the ground state
phase transition in the ferromagnetic model is smeared out by the
longitudinal field in contrast to the antiferromagnetic model for
which the phase transition remains at $h_{z}\neq 0$.

In this paper we study the model (\ref{H}) using the Density
Matrix Renormalization Group (DMRG) technique \cite{white}. This
method allows us to consider the systems up to a few hundred sites
and to determine the transition line with high accuracy. Using the
finite-size estimation of the ground state energy and low-lying
excitations we will show that the model (\ref{H}) on the
transition line is described by the conformal field theory with
the central charge $c=\frac{1}{2}$. Besides, we consider the
behavior of the model in the vicinity of the special points
$h_{x}=\frac{1}{2}$, $h_{z}=0$ and $h_{x}=0$, $h_{z}=1$ and
determine the form of the transition line near these points.

The paper is organized as follows. In Sec.II we will provide a
qualitative physical\ picture of the ground state phase diagram
based on the classical approach. In Sec.III and Sec.IV the
behavior of the system in the vicinity of the special points of
the critical line will be considered. In Sec.V we will present the
DMRG calculation of the critical line. Sec.VI is devoted to the
connection of the model (\ref{H}) with the statistical
quasi-one-dimensional Ising model in the external magnetic field.
In Summary we discuss our results.

\section{The classical approach}

In order to provide a physical picture of the phase diagram of the
model (\ref{H}) we use the classical approximation, when spins are
represented as three-dimensional vectors. The classical ground
state is given by a configuration in which all spin vectors lie in
the $XZ$ plane with the spins on odd and even sites pointing
respectively at angles $\varphi _{1}$ and $-\varphi _{2}$ with
respect to the $X$ axis. The classical energy of this state is
\begin{equation}
E/N =-\frac{1}{4}\sin \varphi _{1}\sin \varphi
_{2}-\frac{h_{x}}{4}(\cos \varphi _{1}+\cos \varphi _{2})
-\frac{h_{z}}{4}(\sin \varphi _{1}-\sin \varphi _{2})
\label{Eclass}
\end{equation}

The angles $\varphi _{1}$ and $\varphi _{2}$, minimizing this
energy are solutions of the following equations
\begin{eqnarray}
\cos \varphi _{1}\sin \varphi _{2}+h_{z}\cos \varphi _{1}-h_{x}\sin \varphi
_{1} &=&0  \nonumber \\
\sin \varphi _{1}\cos \varphi _{2}-h_{z}\cos \varphi _{2}-h_{x}\sin \varphi
_{2} &=&0  \label{eqphi}
\end{eqnarray}

The solution of these equations is simple in the particular cases
$h_{z}=0$
\begin{eqnarray}
\varphi _{1} &=&\varphi _{2}=\arccos (h_{x}),\qquad h_{x}\leq 1
\nonumber
\\
\varphi _{1} &=&\varphi _{2}=0,\qquad h_{x}>1 \label{phihz0}
\end{eqnarray}
and $h_{x}=0$%
\begin{eqnarray}
\varphi _{1} &=&\varphi _{2}=\frac{\pi }{2},\qquad h_{z}\leq 1  \nonumber \\
\varphi _{1} &=&-\varphi _{2}=\frac{\pi }{2},\qquad h_{z}>1
\label{phihx0}
\end{eqnarray}

\begin{figure*}
\includegraphics{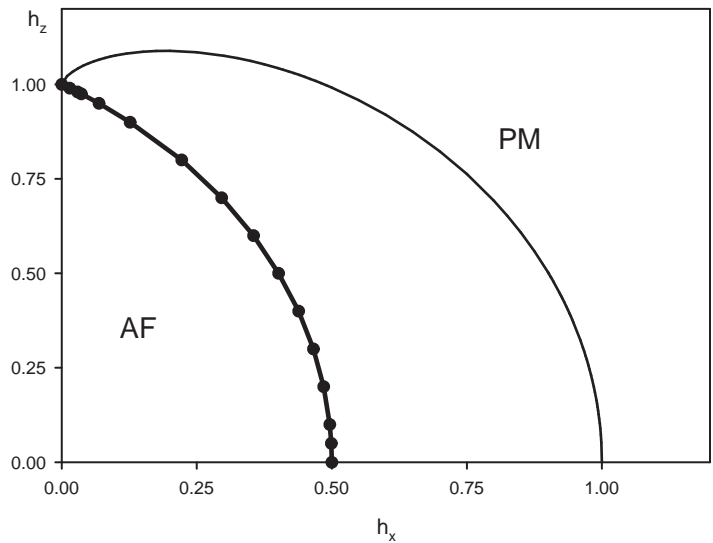}
\caption{\label{fig_1} The ground state phase diagram of the model
(\ref{H}). The critical line between the antiferromagnetic and
paramagnetic states obtained from the DMRG calculation is shown by
thick solid line and that in the classical approximation by thin
solid line.}
\end{figure*}

But in general case, when $h_{x}\neq 0$ and $h_{z}\neq 0$, the
phase diagram is divided on two regions (see Fig.~\ref{fig_1}). In
the paramagnetic region (PM) the energy minimum is given by the
configuration with $\varphi _{1}=-\varphi _{2}=\varphi $, with
$\varphi $ determined by the equation
\begin{equation}
h_{z}\cos \varphi -h_{x}\sin \varphi =\sin \varphi \cos \varphi  \label{PM}
\end{equation}

The antiferromagnetic (AF) region is characterized by non-zero
staggered magnetizations in both $X$ and $Z$ directions
\begin{eqnarray}
S_{2n}^{x}-S_{2n+1}^{x} &=&\sin (\frac{\varphi _{2}-\varphi _{1}}{2})\sin (%
\frac{\varphi _{1}+\varphi _{2}}{2})  \nonumber \\
S_{2n}^{z}-S_{2n+1}^{z} &=&\cos (\frac{\varphi _{2}-\varphi _{1}}{2})\sin (%
\frac{\varphi _{1}+\varphi _{2}}{2})  \label{LROclass}
\end{eqnarray}
with $\varphi _{1}+\varphi _{2}\neq 0$.

Thus, in the AF region the magnetic fields induce a perpendicular
antiferromagnetic long-range order (LRO). In the classical
approach the value $(\varphi _{1}+\varphi _{2})$ plays the role of
the LRO and it vanishes on the transition line determined by the
equations
\begin{eqnarray}
h_{x}\cos \varphi +h_{z}\sin \varphi &=&1  \nonumber \\
h_{z}\cos \varphi -h_{x}\sin \varphi &=&\sin \varphi \cos \varphi
\label{eqtransclass}
\end{eqnarray}
where $\varphi _{1}=-\varphi _{2}=\varphi .$

The solution of Eqs.(\ref{eqtransclass}) gives the transition line
in the explicit form
\begin{equation}
h_{z}=\sqrt{1-h_{x}^{2/3}}(1+h_{x}^{2/3})  \label{trlineclass}
\end{equation}

The classical phase diagram is shown on Fig.~\ref{fig_1}. The
transition (critical) line separates the phase with non-zero AFM
LRO from the phase with uniform magnetization (the paramagnetic
phase).

Of course, the classical approach does not give the correct
description of the phase transition. At first, the quantum
fluctuations shift the critical point $h_{x}=1$ to
$h_{x}=\frac{1}{2}$ at $h_{z}=0$. Secondly, the form of the
critical line is certainly incorrect at $h_{x}\ll 1$. Besides, the
order parameter $(\varphi _{1}+\varphi _{2})$ vanishes on the
critical line with the critical exponent $\frac{1}{2}$, which is
not valid at least in the critical point
$h_{x}=\frac{1}{2},h_{z}=0$. Nevertheless, the fact of the
generation of the staggered magnetizations perpendicular to the
field direction is qualitatively true.

\section{The critical ITF point}

Before the numerical determination of the critical line we
consider the behavior of the model (\ref{H}) in the vicinity of
two special points: $h_{x}=\frac{1}{2},h_{z}=0$ and
$h_{x}=0,h_{z}=1$. This will give us the form of the critical line
near these points.

In the case $h_{z}=0$ the model (\ref{H}) reduces to the exactly
solvable one-dimensional Ising model in the transverse field (the
ITF model). This model is well studied \cite{Pfeuty} and it
belongs to the universality class of the two-dimensional Ising
model. At $h_{x}<\frac{1}{2}$ the ITF model is gapped and there is
non-zero staggered magnetization $\left\langle
(-1)^{n}S_{n}^{z}\right\rangle $. The ground state at
$h_{x}<\frac{1}{2}$ is two-fold degenerated in the thermodynamic
limit. At the point $h_{x}=\frac{1}{2}$ the model becomes gapless
and the staggered magnetization vanishes with the critical
exponent $\frac{1}{8}$. This is the transition point from the
antiferromagnetic state to the paramagnetic gapped state. The gap
vanishes linearly at $h_{x}=\frac{1}{2}$.

Now we consider the ITF model with small longitudinal magnetic
field. For small magnetic field $h_{z}$ we rewrite the Hamiltonian
(\ref{H}) in the form
\begin{eqnarray}
H &=&H_{0}+V  \nonumber \\
H_{0} &=&\sum S_{n}^{z}S_{n+1}^{z}-h_{x}\sum S_{n}^{x}  \nonumber \\
V &=&-h_{z}\sum S_{n}^{z}  \label{HV0}
\end{eqnarray}

After Jordan-Wigner transformation to the Fermi operators
$c_{n}^{+}$, $c_{n} $
\begin{eqnarray}
S_{n}^{z} &=&\frac{c_{n}^{+}+c_{n}}{2}\prod_{j<n}\left(
1-2c_{j}^{+}c_{j}\right)  \nonumber \\
S_{n}^{x} &=&\frac{1}{2}-c_{n}^{+}c_{n}  \label{JW}
\end{eqnarray}
the ITF Hamiltonian $H_{0}$ takes a bilinear form
\begin{equation}
H_{0}=h_{x}\sum \left( c_{n}^{+}c_{n}-\frac{1}{2}\right) +\frac{1}{4}\sum
\left( c_{n}^{+}-c_{n}\right) \left( c_{n+1}^{+}+c_{n+1}\right)
\label{H0fermi}
\end{equation}

The Hamiltonian $H_{0}$ commutes with the 'parity' operator
\begin{equation}
P=\exp \left( i\pi \sum c_{n}^{+}c_{n}\right)  \label{parity}
\end{equation}
because $H_{0}$ can change the number of $c_{n}$ excitation by an
even number only. Therefore, the space of states of $H_{0}$ is
divided on two sectors with odd ($P=-1$) and even ($P=1$) number
of the Fermi particles $c_{n}$.

The Hamiltonian $H_{0}$ is diagonalized exactly \cite{LSM}
\begin{equation}
H_{0}=\sum \varepsilon _{k}\left( \eta _{k}^{+}\eta _{k}-\frac{1}{2}\right)
\label{H0diag}
\end{equation}
with Fermi particles $\eta _{k}$ and spectrum
\begin{equation}
\varepsilon _{k}^{2}=h_{x}^{2}+\frac{1}{4}+h_{x}\cos k  \label{H0spectrum}
\end{equation}
which gives a gap at momentum $\pi $
\begin{equation}
m=\left| h_{x}-\frac{1}{2}\right|  \label{H0gap}
\end{equation}

The gap vanishes at the critical point $h_{x}=1/2$, where the
spectrum becomes
\begin{equation}
\varepsilon _{k}=\cos \frac{k}{2}  \label{Aspectrum}
\end{equation}

Below in this section we will consider the perturbation theory in
$V$ (\ref {HV0}) in the critical point $h_{x}=1/2$.

The transition operator $S^{z}=\sum S_{n}^{z}$ in $V$ (\ref{HV0})
conserves the momentum and changes the 'parity', because as it
follows from Eq.(\ref{JW}) it changes the number of the Fermi
particles $c_{i}$ (and also the number of $\eta _{k}$ particles)
by odd number. Therefore, the non-zero matrix elements with the
transition operator $S^{z}$ have the states with equal momentum
but different parity. The last fact means that the perturbation
theory in $V$ contains only even orders.

The second order correction to the ground state energy is:
\begin{equation}
\delta E_{0}^{(2)}=h_{z}^{2}\sum_{s}\frac{\left\langle 0\right| S^{z}\left|
s\right\rangle \left\langle s\right| S^{z}\left| 0\right\rangle }{E_{0}-E_{s}%
}  \label{dE2}
\end{equation}

The ground state $\left| 0\right\rangle $ has the momentum $q=0$
and zero number of the $\eta _{k}$ particles ($P=1$). Therefore,
the non-zero contribution to the sum in (\ref{dE2}) is given by
the intermediate states $\left| s\right\rangle $ with zero
momentum and the parity $P=-1$. As follows from
Eq.(\ref{Aspectrum}), all states $\left| s\right\rangle $ with
momentum $q=0$ and odd number of the $\eta _{k}$ particles
($P=-1$) have `high' energies $E_{s}-E_{0}\equiv \varepsilon
_{s}\gtrsim 1$. On the contrary, among the states with $q=0$ and
even number of the $\eta _{k}$ particles ($P=1$) there are many
states like $\eta _{\pi -k}^{+}\eta _{-\pi +k}^{+}\left|
0\right\rangle $ with small $k$ having small excitation energies
$\varepsilon _{s}\sim k$ and they can lead to infrared
divergencies.

Hereinafter we consider large but finite systems of length $N$. We
shall study the dependence of dominant contributions to the
perturbation theory on $N$, omitting numerical factors. Using the
fact that $\varepsilon _{s}\gtrsim 1$ one can rewrite
Eq.(\ref{dE2}) as
\begin{equation}
\delta E_{0}^{(2)}\sim -h_{z}^{2}\sum_{s}\left\langle 0\right| S^{z}\left|
s\right\rangle \left\langle s\right| S^{z}\left| 0\right\rangle
=-h_{z}^{2}\left\langle 0\right| \left( S^{z}\right) ^{2}\left|
0\right\rangle  \label{dE22}
\end{equation}

In higher orders of the perturbation series for the ground state
energy each second intermediate state $\left| s\right\rangle $ has
odd number of the Fermi particles ($P=-1$), and, therefore, high
energy $\varepsilon _{s}\geq 1 $. For example, let us consider the
fourth order correction to the ground state energy
\begin{eqnarray}
\delta E_{0}^{(4)} &=&h_{z}^{4}\sum_{s,s^{\prime },s^{\prime \prime }}\frac{%
\left\langle 0\right| S^{z}\left| s\right\rangle \left\langle s\right|
S^{z}\left| s^{\prime }\right\rangle \left\langle s^{\prime }\right|
S^{z}\left| s^{\prime \prime }\right\rangle \left\langle s^{\prime \prime
}\right| S^{z}\left| 0\right\rangle }{\left( E_{0}-E_{s}\right) \left(
E_{0}-E_{s^{\prime }}\right) \left( E_{0}-E_{s^{\prime \prime }}\right) }
\nonumber \\
&&-\delta E_{0}^{(2)}h_{z}^{2}\sum_{s}\frac{\left\langle 0\right|
S^{z}\left| s\right\rangle \left\langle s\right| S^{z}\left| 0\right\rangle
}{\left( E_{0}-E_{s}\right) ^{2}}  \label{dE4}
\end{eqnarray}

All intermediate states $s,s^{\prime },s^{\prime \prime }$ have
momentum $q=0$. The states $s^{\prime }$ have $P=1$ and some of
them have small excitation energies $\varepsilon _{s^{\prime
}}\sim 1/N$, while the states $s$ and $s^{\prime \prime }$ have
$P=-1$ and high energies $\varepsilon _{s}\geq 1$. Therefore, one
can sum over the intermediate states $s$ and $s^{\prime \prime }$,
which reduces Eq.(\ref{dE4}) to
\begin{equation}
\delta E_{0}^{(4)}\sim h_{z}^{4}\sum_{s^{\prime }}\frac{\left\langle
0\right| \left( S^{z}\right) ^{2}\left| s^{\prime }\right\rangle
\left\langle s^{\prime }\right| \left( S^{z}\right) ^{2}\left|
0\right\rangle }{E_{0}-E_{s^{\prime }}}  \label{dE42}
\end{equation}

Now we note, that this expression looks like the second order
correction and Eq.(\ref{dE22}) looks like the first order
correction to the ground state energy with a perturbation
$-h_{z}^{2}\left( S^{z}\right) ^{2}$. In a similar way one can sum
over all intermediate states with $q=0$ and $P=-1$ in all orders
of perturbation series. As a result, we arrive at the perturbation
theory with the effective perturbation
\begin{equation}
V_{1}=-h_{z}^{2}\left( S^{z}\right)
^{2}=-h_{z}^{2}\sum_{n,m}S_{n}^{z}S_{m}^{z}  \label{V1}
\end{equation}

We note, that the perturbation theory with the perturbation
$V_{1}$ coincides with the original perturbation theory
(\ref{HV0}) in a sense, that both perturbation series have the
same order of divergencies (or power of $N$) at each order in
$h_{z}$. But numerical factors at each order in $h_{z}$ can be
different.

The perturbation $V_{1}$ commutes with the parity operator
(\ref{parity}) and conserves the momentum. Therefore, the
perturbation series in $V_{1}$ contains the intermediate states
with $q=0$ and $P=1$ only, and some of these states have small
excitation energies $\varepsilon _{s}\sim 1/N$. These states give
the most divergent contribution to the perturbation series in
$V_{1}$ and further we shall take into account these states only.

Now we need to estimate the matrix elements of the operator
$\left( S^{z}\right) ^{2}$. The behavior of the correlation
function $\left\langle S_{n}^{z}S_{m}^{z}\right\rangle $ in the
ground state and in the low-lying states with excitation energies
$\varepsilon _{s}\sim 1/N$ on large distances is known
\cite{McCoy}
\begin{equation}
\left\langle S_{n}^{z}S_{m}^{z}\right\rangle \sim \frac{(-1)^{n-m}}{\left|
n-m\right| ^{1/4}}  \label{ZZcorr}
\end{equation}
and, therefore, due to oscillation of the correlator $\left\langle
S_{n}^{z}S_{m}^{z}\right\rangle $ the sum over $n$ and $m$ can be
estimated as
\begin{equation}
\sum_{n,m}\left\langle S_{n}^{z}S_{m}^{z}\right\rangle =\frac{N}{4}%
+2N\sum_{n>1}\left\langle S_{1}^{z}S_{n}^{z}\right\rangle \simeq 0.07465(1)N
\label{V1diag}
\end{equation}
where the constant was found from extrapolation of the exact
results for finite chains with $N=6,\ldots 14$.

The non-diagonal matrix elements of the operator $\left(
S^{z}\right) ^{2}$ with two different low-lying states was also
calculated numerically. It was found that the only non-zero matrix
elements (in the thermodynamic limit) are given by the states
$s,s^{\prime }$ differing by two $\eta _{k}$ particles, like
$\left| s^{\prime }\right\rangle =\eta _{k}^{+}\eta
_{-k}^{+}\left| s\right\rangle $. All such pairs of states give
the same value for the matrix element
\begin{equation}
\left\langle s\right| \left( S^{z}\right) ^{2}\left| s^{\prime
}\right\rangle =0.3108(1)  \label{V1nodiag}
\end{equation}
while all other matrix elements exponentially drop with $N$.

Now we estimate all terms of perturbation series in $h_{z}$.
According to Eq.(\ref{V1diag}) the second order correction
(\ref{dE22}) is proportional to $h_{z}^{2}N$. The factor at
$h_{z}^{2}N$ was found numerically by exact diagonalization of
finite systems and the following calculation of the sum
(\ref{dE2}). The data for $N=6,\ldots 14$ are well extrapolated
and give
\begin{equation}
\delta E_{0}^{(2)}=-0.07060(5)h_{z}^{2}N  \label{dE23}
\end{equation}
\qquad

According to Eq.(\ref{dE23}) the zero-field susceptibility is
$\chi _{z}=0.1412(1)$. This result is in a perfect agreement with
the value $\chi _{z}=0.14118\ldots $ obtained analytically in
\cite{muller}.

Using Eq.(\ref{V1nodiag}) we find that the non-zero contribution
to the fourth order correction to the ground state energy
(\ref{dE42}) is given by the states $\left| s\right\rangle =\eta
_{k}^{+}\eta _{-k}^{+}\left| 0\right\rangle $ with the excitation
energy $\varepsilon _{s}=2\cos \frac{k}{2}$. Thus, because of
small excitation energies at $k\sim \pi $ in denominator in
(\ref{dE42}) the fourth order correction to the ground state
energy turns out to be proportional to $N$ (we omit here
logarithmic corrections)
\begin{equation}
\delta E_{0}^{(4)}\sim h_{z}^{4}\sum_{s}\frac{1}{\varepsilon _{s}}\sim
h_{z}^{4}N  \label{dE43}
\end{equation}

The $m$-th order in $V_{1}$ is proportional to $h_{z}^{2m}$. The
denominator of the $m$-th order contains $(m-1)$ small excitation
energies $\sim 1/N$ and all matrix elements in the numerator are
of the order of unity. Therefore, the $m$-th order in $V_{1}$
diverges as $h_{z}^{2m}N^{m-1}$. But this is not valid for odd
orders in $V_{1}$. The analysis shows that all odd orders in
$V_{1}$ diverge as $h_{z}^{2m}N^{m-2}$. For example, the third
order correction in $V_{1}$ (the sixth order in $h_{z}$) for the
ground state energy does not diverge:
\begin{equation}
\delta E_{0}^{(6)}\sim h_{z}^{6}N  \label{dE6}
\end{equation}

So, the odd orders in $V_{1}$ give the next order corrections to
the ground state energy and we omit it below.

Summarize all above, we arrive at the perturbation series in the
form
\begin{equation}
\delta E_{0}=-a_{0}h_{z}^{2}N-h_{z}^{4}N\sum_{n=0}^{\infty }b_{n}\left(
h_{z}^{2}N\right) ^{n}  \label{dE0ser}
\end{equation}
with $a_{0}=0.07059\ldots $ (Eq.(\ref{dE23})) and unknown
constants $b_{n}$. One can see that the first divergence appears
only in the eight-th order in $h_{z}$, and it is very difficult to
observe it numerically. The exact numerical calculations up to
$N=14$ of the second, the fourth and the sixth order corrections
in $h_{z}$ confirm the form of the series (\ref{dE0ser}).

The sum in Eq.(\ref{dE0ser}) forms the scaling function
$f_{0}\left( x\right) $ of the scaling parameter $x=h_{z}^{2}N$.
Thus, the ground state energy takes the form
\begin{equation}
\delta E_{0}=-a_{0}h_{z}^{2}N-h_{z}^{4}Nf_{0}\left( x\right)  \label{dE0}
\end{equation}

Since the ground state energy is proportional to $N$, the scaling
function $f_{0}\left( x\right) $ has finite thermodynamic limit at
$x\rightarrow \infty $. Thus, the leading term of the perturbation
theory for the ground state energy is given by the second order
and the divergent part of the perturbation theory gives the
correction $\sim h_{z}^{4}$.

The perturbation series for low-lying states has the same form as
in Eq.(\ref {dE0}), but each low-lying state has its own scaling
function. Therefore, for the first excited state (with momentum
$q=\pi $) one has
\begin{equation}
\delta E_{\pi }=-a_{\pi }h_{z}^{2}N-h_{z}^{4}Nf_{\pi }\left( x\right)
\label{dEpi}
\end{equation}
\qquad

So, the mass gap $m=\delta E_{\pi }-\delta E_{0}$ appears as:
\begin{equation}
m=a_{m}h_{z}^{2}+g\left( x\right) h_{z}^{2}  \label{mser}
\end{equation}
where $a_{m}=(a_{0}-a_{\pi })N$ and the scaling function $g\left(
x\right) =x\left( f_{0}\left( x\right) -f_{\pi }\left( x\right)
\right) $. Since the gap is finite, the scaling function $g\left(
x\right) $ in the thermodynamic limit at $x\rightarrow \infty $
must tend to some finite limit
\begin{equation}
m=a_{m}h_{z}^{2}+g\left( \infty \right) h_{z}^{2}  \label{m}
\end{equation}

From the last equation we see that the gap is proportional to
$h_{z}^{2}$, but the factor at $h_{z}^{2}$ is given not only by
the second order correction $a_{m}$ but by all collected divergent
orders of the perturbation series. The numerical estimation of the
second order correction to the gap gives
\begin{equation}
a_{m}=0.1875(2)  \label{m2}
\end{equation}

\begin{figure*}
\includegraphics{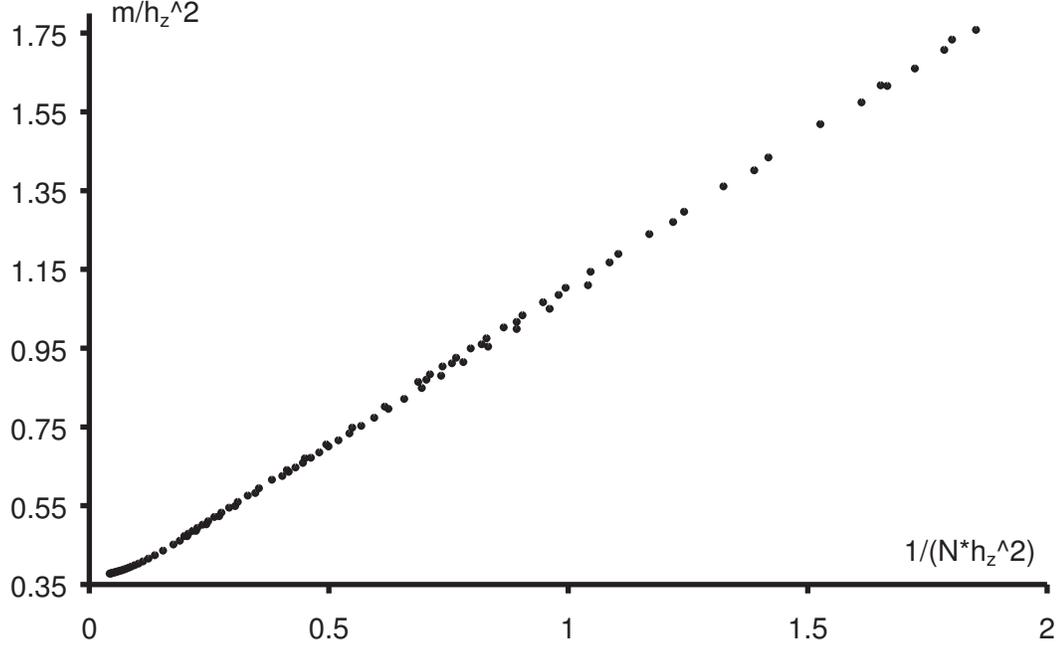}
\caption{\label{fig_2} Scaled mass gap $\frac{m}{h_z^2}$ for
$h_{x}=\frac 12$ with various values of $N$ and $h_z$ as a
function of the reciprocal scaled parameter $\frac{1}{Nh_z^2}$.}
\end{figure*}

In order to find the factor at $h_{z}^{2}$ in mass gap we
performed DMRG calculations (for details see Sec.V) of the model
(\ref{HV0}) for $h_{z}=0,\ldots 0.3$ and $N=20,...300$. The
dependence of the gap on the scaling parameter $x=h_{z}^{2}N$ is
shown on Fig.~\ref{fig_2}. One can see that the points with
different $N$ and $h_{z}$ lie perfectly on one curve. It
definitely manifests that the scaling parameter is $h_{z}^{2}N$.
The DMRG calculation data give the mass gap
\begin{equation}
m=0.37(1)h_{z}^{2}  \label{mvalue}
\end{equation}

From this equation follows that the second order correction to the
mass gap\ (\ref{m2}) gives approximately one half of a total gap,
and another half is collected by all other divergent orders
contained in the scaling function $g\left( x\right) $.

We note, that in contrast to Eq.(\ref{mvalue}) for the
ferromagnetic ITF\ model the mass gap at the critical point
$h_{x}=\frac{1}{2}$ is proportional to $h_{z}^{8/15}$
\cite{zam},\cite{fateev}.

\subsection{Mapping to the effective model}

Let us return to the estimation of the diagonal element of the
operator $V_{1}$ (\ref{V1diag}). Using the fact of the oscillation
of the correlator (\ref{ZZcorr}) in the ground state and in the
low-lying states we can rewrite the sum in (\ref{V1diag})
approximately as a one half of the first term:
\begin{equation}
\left\langle \left( S^{z}\right) ^{2}\right\rangle =\frac{N}{4}%
+2\sum_{n<m}\left\langle S_{n}^{z}S_{m}^{z}\right\rangle \simeq \frac{N}{4}%
+\sum_{n}\left\langle S_{n}^{z}S_{n+1}^{z}\right\rangle  \label{mapdiag}
\end{equation}

The last equation suggests that the perturbation $V_{1}$ can be
reduced to the operator $V_{2}$
\begin{equation}
V_{1}\rightarrow V_{2}=-ah_{z}^{2}N-bh_{z}^{2}\sum S_{n}^{z}S_{n+1}^{z}
\label{V2}
\end{equation}
with some constants $a,b$, which we will define later. In order to
verify this assumption we compared the matrix elements of the
operator $V_{2}$ with those of $V_{1}$. We have found that the
dependence of the matrix elements of the operator $V_{2}$ on $N$
is in a full accord with Eqs.(\ref{V1diag}) and (\ref{V1nodiag}).
That is, the diagonal matrix elements of $V_{2}$ are proportional
to $N$ and the only non-zero (non-diagonal) matrix elements have
the pairs of states differing by two $\eta _{k}$ particles
\begin{equation}
\left\langle s\right| S_{n}^{z}S_{n+1}^{z}\left| s^{\prime }\right\rangle =%
\frac{1}{2}  \label{V2nodiag}
\end{equation}

Thus, we arrive at the effective Hamiltonian
\begin{equation}
H_{{\rm eff}}=-ah_{z}^{2}N+(1-bh_{z}^{2})\sum S_{n}^{z}S_{n+1}^{z}-\frac{1}{2%
}\sum S_{n}^{x}  \label{Hmap}
\end{equation}

Again, the original model (\ref{HV0}) is equivalent to the
effective model (\ref{Hmap}) in a sense, that the perturbation
series for both models have the same order of divergencies (or
power of $N$) at each order of $h_{z}$. But numerical factors at
each order of $h_{z}$ can be different.

The advantage of this mapping lies in the fact that the effective
model (\ref {Hmap}) is of the ITF type and, therefore, is exactly
solvable one. The spectrum of $H_{{\rm eff}}$ is
\begin{equation}
\varepsilon _{k}^{2}=\frac{1+J^{2}}{4}+\frac{J}{2}\cos k
\label{Hmapspectrum}
\end{equation}
with $J=1-bh_{z}^{2}$. So, according to the spectrum of the
effective model (\ref{Hmapspectrum})\ the magnetic field $h_{z}$
produces the gap in the spectrum
\begin{equation}
m=\frac{b}{2}h_{z}^{2}  \label{gapmap}
\end{equation}
\qquad

As was noted above, apart from numerical factors the effective
model (\ref {Hmap}) has the same form of perturbation series in
$h_{z}$ as the original model (\ref{HV0}). Therefore, we can check
the form of the perturbation theory (\ref{dE0ser}) by studying the
perturbation series in $h_{z}$ of the exactly solvable model
(\ref{Hmap}).

For large but finite systems the ground state energy (with $q=0$)
and the first excited state (with $q=\pi $) of the Hamiltonian
(\ref{Hmap}) are known exactly \cite{hamer}:
\begin{eqnarray}
E_{0} &=&-ah_{z}^{2}N-\frac{1}{4}\sum_{j=1}^{N}\sqrt{g^{2}+4(1-g)\sin
^{2}\left( \frac{\pi j}{N}-\frac{\pi }{2N}\right) }  \nonumber \\
E_{\pi } &=&-ah_{z}^{2}N+\frac{g}{4}-\frac{1}{4}\sum_{j=1}^{N-1}\sqrt{%
g^{2}+4(1-g)\sin ^{2}\left( \frac{\pi j}{N}\right) }  \label{Hmapenergy}
\end{eqnarray}
where $g=bh_{z}^{2}$.

The perturbation series for the ground state energy and the gap
can be easily found by formal expansion of these expressions in
small parameter $g$. This results in the scaling form of
perturbation series for the mass gap:
\begin{equation}
m=\frac{g}{4}+g^{2}N\sum_{n=0}^{\infty }d_{n}\left( gN\right) ^{2n}
\label{Hmapser}
\end{equation}
where constants $d_{n}$ determine the scaling function $G(x)$ of a scaling
parameter $x=gN$
\begin{equation}
m=\frac{g}{4}+gG(x)  \label{mapscaling}
\end{equation}

The scaling function $G(x)$ was found in \cite{hamer}. In the
thermodynamic limit $x\rightarrow \infty $ the function
$G(x)\rightarrow 1/4$, resulting in the gap (\ref{gapmap}). One
can see that the scaling parameter and the form of the
perturbation series (\ref{Hmapser}) coincides with
Eq.(\ref{mser}).

Now we define the constants $a$ and $b$ in Eq.(\ref{Hmap}), so
that the second order corrections to the ground state energy and
the gap of the effective model (\ref{Hmap}) numerically coincide
with Eqs.(\ref{dE23}) and (\ref {m2}). This leads to the values
\begin{eqnarray}
a &=&0.190(1)  \nonumber \\
b &=&0.750(1)  \label{ab}
\end{eqnarray}

The mass gap of the effective model (\ref{Hmap}) with constants
$a$ and $b$ defined above is
\begin{equation}
m=0.375(1)h_{z}^{2}  \label{Hgap}
\end{equation}

Surprisingly, the value of the mass gap of the effective model
turns out to be very close to one found numerically for the
original model (Eq.(\ref {mvalue})). It means, that the gap
scaling function $G(x)$ of the effective model (\ref{Hmap})\ has
the same thermodynamic limit as that for the model (\ref{HV0}).
Moreover, we have calculated numerically the\ fourth order
corrections in $h_{z}$ to the ground state energy and the gap for
the models (\ref{HV0}) and (\ref{Hmap}) and found perfect
numerical agreement (the second order corrections in $h_{z}$ for
both models coincide by the definition (\ref{ab})). Therefore, we
expect that the mapping of the model (\ref{HV0}) at $h_{x}=1/2$ to
the model (\ref{Hmap}) with constants $a$ and $b$ defined in
(\ref{ab}) is exact for low-lying excitations in the thermodynamic
limit.

The mapping of the model (\ref{HV0}) to the effective ITF model
can be extended to the case $h_{x}\neq 1/2$. For this case the
effective Hamiltonian becomes
\begin{equation}
H_{\rm eff} =-a(h_{x})h_{z}^{2}N +\left(
1-b(h_{x})h_{z}^{2}\right) \sum S_{n}^{z}S_{n+1}^{z}-h_{x}\sum
S_{n}^{x}  \label{Hmaphx}
\end{equation}
where now the constants $a,b$ are the functions of $h_{x}$. The
functions $a(h_{x})$ and $b(h_{x})$ are defined in such a way that
the second order corrections to the ground state energy and the
gap for the effective model (\ref{Hmap}) coincide with those for
the original model (\ref{HV0}).

Using the effective Hamiltonian (\ref{Hmaphx}) one can calculate
the susceptibility $\chi _{z}(h_{x})$. In the vicinity of the
critical point $h_{x}=1/2$ it has a logarithmic singularity
\begin{equation}
\chi _{z}(h_{x})=\chi _{zc}-\frac{b}{\pi }\left( h_{x}-\frac{1}{2}\right)
\log \left( h_{x}-\frac{1}{2}\right)  \label{chi}
\end{equation}
with $\chi _{zc}=0.1412$ from Eq.(\ref{dE23}) and $b=0.75$
(Eq.(\ref{ab})).

\section{The multicritical point}

Another exactly solvable limit of the model (\ref{H}) is the case
$h_{x}=0$. The model at $h_{x}=0$ is the classical one. At
$h_{z}<1$ its ground state is the antiferromagnet. At $h_{z}=1$
the first order phase transition to the ferromagnetic ground state
occurs. This is so-called multicritical point. The Hamiltonian
(\ref{H}) at $h_{z}=1$ has a form
\begin{eqnarray}
H &=&H_{0}+V  \nonumber \\
H_{0} &=&-\frac{N}{4}+\sum \left( S_{n}^{z}-\frac{1}{2}\right) \left(
S_{n+1}^{z}-\frac{1}{2}\right)  \nonumber \\
V &=& -\frac{h_{x}}{2}\left( S^{+}+S^{-}\right)  \label{Hhz1}
\end{eqnarray}
where $S^{\pm }=\sum S_{n}^{\pm }$.

The ground state of $H_{0}$ is macroscopic degenerate: all spin
configurations, excluding those with two neighbor spins pointing
down, have the same energy $-\frac{N}{4}$. The number of these
states is $(\frac{1+\sqrt{5}}{2})^{N}$ \cite{domb}. The transverse
field $h_{x}$ lifts the degeneracy. The exact calculation of the
first order correction in $h_{x}$ for $N\gg 1$ is rather
complicated because it involves the exponentially large number of
degenerate states. We carried out the approximate calculation of
the perturbation theory in $V$ within degenerate manifold using a
simple variational function in the form
\begin{equation}
\Psi =\sum_{m=0}^{N/2}c_{m}\Psi _{m}  \label{psi}
\end{equation}
where $\Psi _{m}$ are the sum (with equal weights) of all
admissible states with $m$ spins down:
\begin{equation}
\Psi _{m}=w_{m}^{-\frac{1}{2}}(S^{-}P)^{m}\Psi _{F}  \label{psim}
\end{equation}

Here $P$ is a projector excluding states with two neighbor down
spins, $\Psi _{F}$ is the ferromagnetic state with all spins up
and the normalization factors $w_{m}$ are
\begin{equation}
w_{m}=\frac{(N-m-1)!N}{m!(N-2m)!}  \label{bm}
\end{equation}

Matrix elements of $V$ with respect to $\Psi _{m}$ are
\begin{equation}
\langle \Psi _{m}V\Psi _{m^{\prime }}\rangle =\frac{h_{x}}{2}(s_{m}\delta
_{m^{\prime },m-1}+s_{m+1}\delta _{m^{\prime },m+1})  \label{Vm1m2}
\end{equation}
where
\begin{equation}
s_{m}=\sqrt{\frac{m(N-2m+2)(N-2m+1)}{N-m}}  \label{sm}
\end{equation}

Coefficients $c_{m}$ in (\ref{psi}) obey the equations
\begin{equation}
(E+\frac{N}{4})c_{m}=\frac{h_{x}}{2}(s_{m+1}c_{m+1}+s_{m}c_{m-1})  \label{cm}
\end{equation}

The quantity $s_{m}$ has a sharp maximum for $N\gg 1$ at $m_{0}=\frac{3-%
\sqrt{5}}{4}N$ and $s_{m_{0}}$ is
\begin{equation}
s_{m_{0}}=\frac{N}{2}(\sqrt{5}-1)\sqrt{\sqrt{5}-2}  \label{sm0}
\end{equation}

The ground state energy in the thermodynamic limit is defined as a
lowest eigenvalue of Eqs.(\ref{cm}), which is
\begin{equation}
E_{0}=-\frac{N}{4}-h_{x}s_{m_{0}}=-\frac{N}{4}-0.30028Nh_{x}  \label{Ehx0}
\end{equation}

In the frame of variational function approach (\ref{psi}) the
ground state magnetizations $\langle S_{n}^{z}\rangle $ and
$\langle S_{n}^{x}\rangle $ are
\begin{eqnarray}
\langle S_{n}^{z}\rangle &=&\frac{1}{2}-\frac{m_{0}}{2N}=0.309  \nonumber \\
\langle S_{n}^{x}\rangle &=&\frac{s_{m_{0}}}{N}=0.300028  \label{magnhx0}
\end{eqnarray}

The function (\ref{psi}) has a momentum $q=0$. To calculate the
spectrum in the first order in $h_{x}$ it is necessary to choose
the variational function of the type (\ref{psi}) with momentum
$q$. We omit here rather cumbersome calculations, which shows that
the mass gap corresponds to $q=\pi $ and
\begin{equation}
m=h_{x}  \label{mvar}
\end{equation}

The numerical diagonalization of finite cyclic systems shows very
rapid exponential convergence to the thermodynamic limit.\ The
extrapolated values for the ground state energy and the mass gap
(at $q=\pi $) for $h_{x}\ll 1$ are
\begin{eqnarray}
\frac{E_{0}}{N} &=&-\frac{1}{4}-0.3017(1)h_{x}  \nonumber \\
m &=&0.4841(1)h_{x}  \label{Emhx0}
\end{eqnarray}

Comparison of these results with Eq.(\ref{Ehx0}) shows that the
variational ground state energy differs from the `exact' one
within $0.4\%$. As follows from Eq.(\ref{Emhx0}) the mass gap in
the multicritical point opens linearly with $h_{x}$, which is
correctly described by the variational approach (\ref {mvar}).

\section{The critical line}

In general, the critical line at $h_{x}\neq 0$ and $h_{z}\neq 0$
can not be found exactly. To obtain it we used the DMRG technique
\cite{white}. We have performed\ DMRG calculations using both the
infinite-size and the finite-size DMRG algorithms. We calculated
the ground state energy $E_{0}(N)$ and two lowest excitations
$m_{1}(N)=E_{1}(N)-E_{0}(N)$ and $m_{2}(N)=E_{2}(N)-E_{0}(N)$.

In order to check the accuracy of the DMRG method we compared the
obtained results with the exact ones for the ITF model. We used
the infinite-size algorithm and open boundary conditions. The
dependence of the results on the number of retained states $s$ in
the DMRG computation and on a number $N_{RG} $ of DMRG steps
($N=2N_{RG}+2$) has been investigated. We have found that the
calculation with $s=25$ gives satisfactory accuracy up to $N=300$.
For example, relative errors in the ground state energy and in the
mass gap at $h_{x}=\frac{1}{2}$ are $\frac{\Delta
E_{0}}{E_{0}}=10^{-9}(10^{-7})$ and $\frac{\Delta
m}{m}=10^{-7}(10^{-5})$ for $N=100(300)$. The accuracy becomes
better when the value $\left| h_{x}-\frac{1}{2}\right| $ is
increased.

\begin{figure*}
\includegraphics{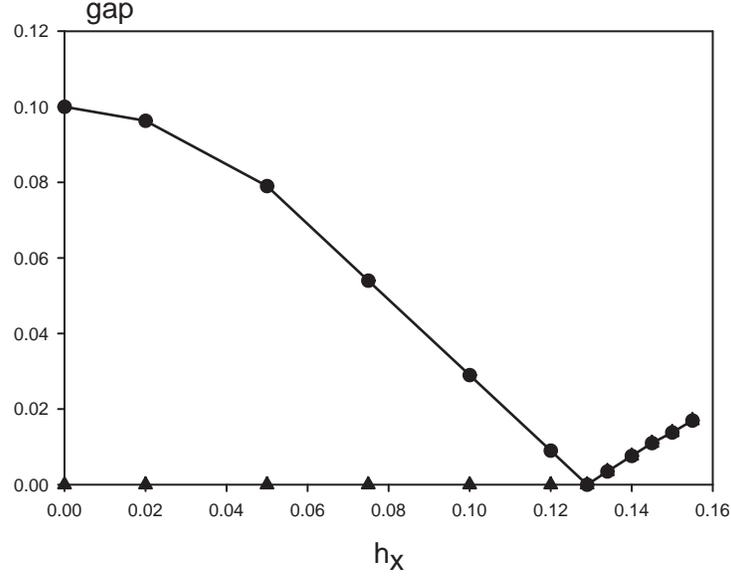}
\caption{\label{fig_3} The mass gaps $m_{1}$ (triangles) and
$m_{2}$ (circles) as functions of the transverse field $h_{x} $ at
$h_{z}=0.9$ from the DMRG calculations extrapolated to the
thermodynamic limit.}
\end{figure*}

The critical field $h_{x{\rm c}}(h_{z})$ at a fixed value $h_{z}$
($0<h_{z}<1 $) is determined by vanishing of the gaps $m_{1}$ and
$m_{2}$. Below the critical field the mass gap
$m_{1}(N)\rightarrow 0$ exponentially with $N$. This behavior
confirms the fact that the ground state is doubly degenerate in
the thermodynamic limit at $h_{x}<h_{x{\rm c}}(h_{z})$. The true
mass gap in this region is defined by the value of $m_{2} $. The
typical behavior of the gaps $m_{1}$and $m_{2}$ extrapolated to
the thermodynamic limit is shown on Fig.~\ref{fig_3}. We note that
at $h_{x}>h_{x{\rm c}}(h_{z})$ the gaps $m_{1}$and $m_{2}$
coincide at $N\rightarrow \infty $.

The critical line obtained from the DMRG results extrapolated to
the thermodynamic limit is shown on Fig.~\ref{fig_1}. In the plane
$(h_{x},h_{z})$ it connects two limiting critical points studied
in previous sections: the critical point
($h_{x}=\frac{1}{2},h_{z}=0$) of the ITF model and the
multicritical point ($h_{x}=0,h_{z}=1$). Now we discuss the
properties of the model (\ref{H}) near the critical line.

The ITF model in the critical point is described by a conformal
field theory with a central charge $c=\frac{1}{2}$. We expect that
the ITF model is generic for the model (\ref{H}) on the whole
critical line (except the multicritical point $h_{x}=0,h_{z}=1$).
To verify this suggestion we estimated the value of $c$ on the
critical line. For this we used well known fact \cite{cardy} that
for conformal invariant model with periodic boundary conditions
the central charge appears at $1/N$ correction to the ground state
energy
\begin{equation}
E_{0}=e_{\infty }N-\frac{\pi c\upsilon }{6N}  \label{EgsCFT}
\end{equation}
where $e_{\infty }$ is the ground state energy per site in the
thermodynamic limit and $\upsilon $ is the sound velocity.

At first, we calculated the sound velocity $\upsilon $ as (similar
to the spectrum of the ITF model (\ref{Aspectrum}) the sound
velocity on the whole critical line is determined at $k=\pi $)
\begin{equation}
\upsilon (N)=\frac{N}{2\pi }\left[ E(\pi -\frac{2\pi }{N})-E(\pi )\right]
\label{vCFT}
\end{equation}

\begin{figure*}
\includegraphics{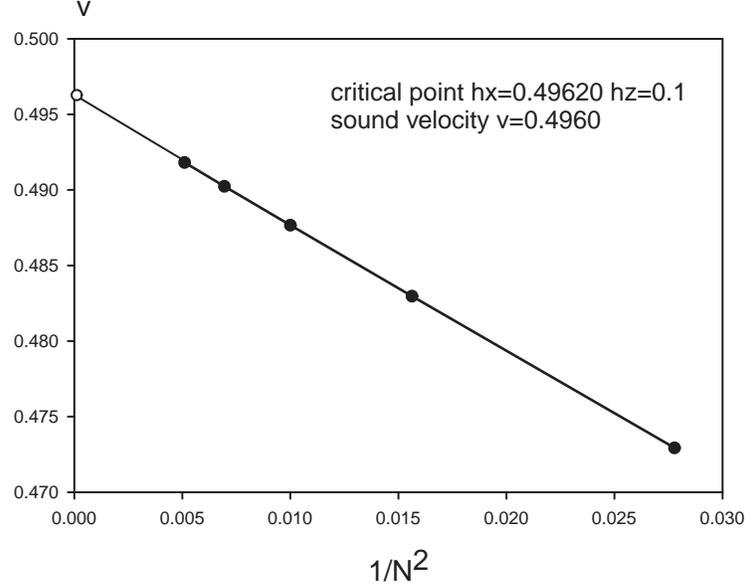}
\caption{\label{fig_4} The extrapolation procedure of finite size
dependence of the sound velocity in the critical point
$h_{x}=0.4962,h_{z}=0.1$. ($v=0.4960$).}
\end{figure*}

\begin{figure*}
\includegraphics{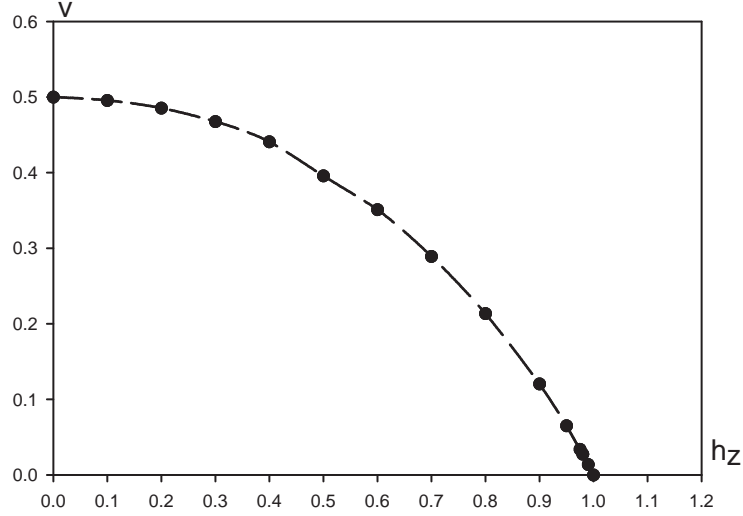}
\caption{\label{fig_5} The dependence of the sound velocity on
$h_{z}$ along the critical line.}
\end{figure*}

We carried out these calculations using the numerical
diagonalization of the Hamiltonian (\ref{H}) with the periodic
boundary condition for $N\leq 14$. The size extrapolation is
carried out by formula $\upsilon (N)=\upsilon +aN^{-2}$. The
example of the extrapolation procedure is shown on
Fig.~\ref{fig_4}. The dependence of $\upsilon $ on $h_{z}$ along
the critical line is shown on Fig.~\ref{fig_5}. After that, the
central charge $c$ has been calculated using Eq.(\ref {EgsCFT}).
The size extrapolation shows that the central charge is
$c=0.500(1)$ for all calculated critical points.

The critical exponents $\nu $ of the ground state correlation
function $\langle \widehat{O}_{1}\widehat{O}_{R}\rangle \sim
R^{-\nu }(R\gg 1)$ is defined by the scaling dimension $X$ of the
operator $\widehat{O}$ and $\nu =2X$ \cite{book}. The scaling
dimension is related to the finite-size correction of the lowest
excited eigenstate which can be reached from the ground state by
the operator $\widehat{O}$. The finite-size correction to this
excitation energy is
\begin{equation}
\Delta E_{i}=E_{i}-E_{0}=\frac{2\pi \upsilon }{N}X_{i}+o(\frac{1}{N})
\label{deCFT}
\end{equation}

For the ITF model at $h_{x}=\frac{1}{2}$ the sound velocity is
$\upsilon = \frac{1}{2}$. The ground state has a momentum $q=0$
and the lowest excitation energy in the sector with momentum $q=0$
is
\begin{equation}
E(q=0)-E_{0}=2\sin \frac{\pi }{2N}  \label{dE0ex}
\end{equation}
and in the sector with $q=\pi $ is
\begin{equation}
E(q=\pi )-E_{0}=\frac{1}{2}\tan \frac{\pi }{4N}  \label{dEpiex}
\end{equation}

Thus, the scaling dimensions are $X_{0}=1$ and $X_{\pi
}=\frac{1}{8}$. The corresponding associated operators are
\begin{eqnarray}
\widehat{O}_{q=0} &=&\sum_{n}S_{n}^{x},  \nonumber \\
\widehat{O}_{q=\pi } &=&\sum_{n}(-1)^{n}S_{n}^{z}  \label{Op}
\end{eqnarray}
which is in accord with the well known results \cite{McCoy} for
the asymptotic of the correlation functions
\begin{eqnarray}
\langle S_{0}^{x}S_{R}^{x}\rangle -\langle S_{0}^{x}\rangle ^{2} &\sim
&R^{-2}  \nonumber \\
\langle S_{0}^{z}S_{R}^{z}\rangle &\sim &\frac{(-1)^{R}}{R^{1/4}}
\label{ITFcorr}
\end{eqnarray}

\begin{figure*}
\includegraphics{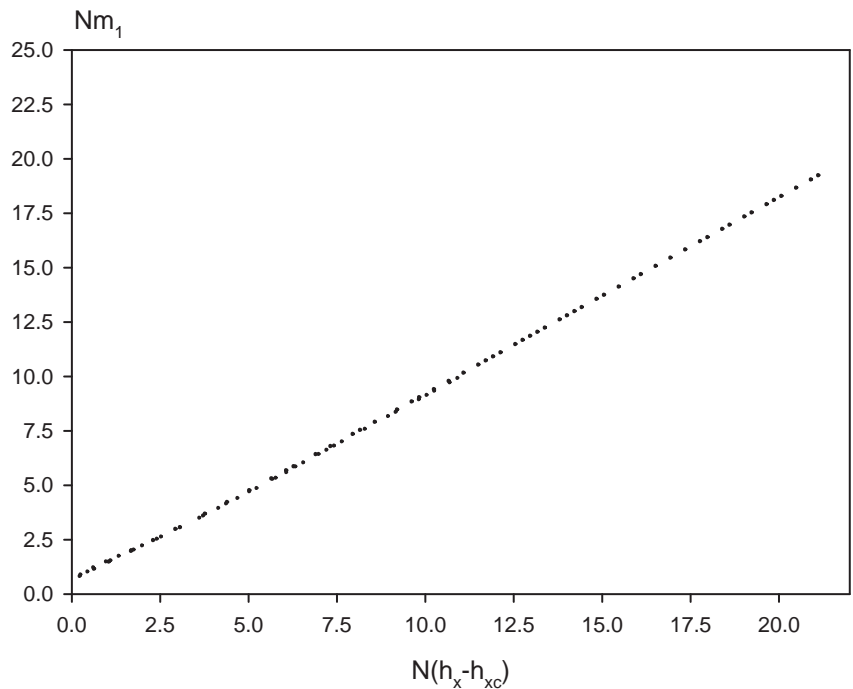}
\caption{\label{fig_6} Scaled mass gap $Nm_{1}$ for $h_{z}=0.5$
with various values of $N$ plotted as a function of the scaled
field $Nh_{x}$.}
\end{figure*}

Using Eq.(\ref{deCFT}) we numerically checked that the scaling
dimensions related to the finite-size corrections of the lowest
excited eigenstates on the critical line remain as in ITF model
$X_{0}=1$ and $X_{\pi }=\frac{1}{8} $. Therefore, we conclude that
the model (\ref{H}) on the critical line belongs to the
universality class of the ITF model. This means that in accord
with the prediction of the classical approach to the left of the
critical line the staggered magnetizations $\langle (-1)^n
S_{n}^{z} \rangle$ and $\langle (-1)^n S_{n}^{x} \rangle$ exist.
But in contrast to the classical approach they vanish on the
critical line with the critical exponent $\frac{1}{8}$. The mass
gap is closed on the critical line and the critical exponent for
the gap is equal to unity, i.e. the scaling behavior of the gap
near the critical line is linear. For the particular case
$h_{z}=0.5$, this is illustrated on Fig.~\ref{fig_6}. where the
scaling plot for the scaled mass $Nm_{1}$ with $N(h_{x}-h_{x{\rm
c}})$ is shown.

The behavior of the critical line near the ITF point
$h_{x}=1/2,h_{z}=0$ can be found from the following consideration.
As it was established above in the vicinity of the critical line
the gap is proportional to deviation from the line. This is valid
for any direction of deviation except the direction at a tangent
to the critical line. In the vicinity of the ITF critical point
for fixed $h_{z}\ll 1$ according to Eq.(\ref{H0gap}) the gap is
$m=\left| h_{x}-h_{x{\rm c}}(h_{z})\right| $. On the other hand,
the gap is given by Eq.(\ref{mvalue}). Combining these two
expressions for the gap on the line $h_{x}=1/2$, we obtain
equation for the critical line in the vicinity of the point
$h_{x}=1/2,h_{z}=0$ as
\begin{equation}
h_{x{\rm c}}(h_{z})=\frac{1}{2}-0.37h_{z}^{2}  \label{linehz0}
\end{equation}

This expression for the critical line is in a very good agreement
with our numerical estimations up to $h_{z}\sim 0.5$.

In the vicinity of the multicritical point the form of the
critical line can be found in a similar way. At $h_{z}=1$ and
$h_{x}\ll 1$ the mass gap is proportional to $h_{x}$ (see
Eq.(\ref{Emhx0})). On the other hand, for fixed $h_{x}\ll 1$ this
gap is proportional to a deviation from the critical line $m\sim
\left( h_z - h_{z{\rm c}}\right) $. Therefore, the critical line
near the multicritical point behaves as $h_{x{\rm c}}=A(1-h_{z})$.
The numerical coefficient $A$ has been found from the DMRG
calculation of the critical line in the vicinity of the
multicritical point. As a result, the form of the critical line at
$h_{x}\ll 1$ is
\begin{equation}
h_{x{\rm c}}(h_{z})=1.50\left( 1-h_{z}\right) +O\left( \left(
1-h_{z}\right) ^{2}\right)  \label{linehx0}
\end{equation}

This form differs from that given in \cite{sen}, where $h_{x{\rm
c}}(h_{z})\sim \left( 1-h_{z}\right) ^{0.75}$.

\section{Relation with the two-dimensional Ising model}

It is well-known that the critical properties of the
two-dimensional Ising model are equivalent to those of the
one-dimensional ITF model \cite{kogut}. Using the formalism of
Ref.\cite{kogut} one can show that the quantum Hamiltonian
(\ref{H}) is related to the transfer matrix of the strong
anisotropic (quasi-one-dimensional) Ising model in the uniform
magnetic field. This model describes ferromagnetic Ising chains
weakly antiferromagnetically coupled with each other. The
Hamiltonian of the model is
\begin{equation}
H=-J_{1}\sum \sigma _{n,m}\sigma _{n+1,m}+J_{2}\sum \sigma _{n,m}\sigma
_{n,m+1}-h\sum \sigma _{n,m}  \label{H2d}
\end{equation}
where $\sigma _{n,m}=\pm 1$, $J_{1}\gg J_{2}>0$.

The relation between the model (\ref{H}) and the model (\ref{H2d})
are given by
\begin{eqnarray}
2h_{x}\beta J_{2} &=&e^{-2\beta J_{1}}\ll 1  \nonumber \\
h &=&2h_{z}J_{2}  \label{1d-2d}
\end{eqnarray}
with $\beta =1/kT$.

The equivalence of these two models means that the model
(\ref{H2d}) undergoes the phase transition from the ordered (at
$T<T_{c}$) to the disordered (at $T>T_{c}$) phase. The critical
line $T_{c}(h)$ of the quasi-one dimensional model are related to
the critical line $h_{x{\rm c}}(h_{z})$ of the model (\ref{H}). By
use of mapping (\ref{1d-2d}) we obtain that the critical
temperature $T_{c}(h)$ is
\begin{equation}
kT_{c}(h)=\frac{2J_{1}}{\log \left( \frac{J_{1}}{J_{2}h_{x}(h/2J_{2})}%
\right) }  \label{Tc(h)}
\end{equation}

In particular, at $h\rightarrow 0$
\begin{equation}
kT_{c}=\frac{2J_{1}}{\log \left( \frac{2J_{1}}{J_{2}}\right) }  \label{Tch0}
\end{equation}
and $T_{c}\rightarrow 0$ when $h\rightarrow 2J_{2}$ as
\begin{equation}
kT_{c}=\frac{2J_{1}}{\log \left( \frac{J_{1}}{2J_{2}-h}\right) }
\label{Tch2}
\end{equation}

The order parameter $\frac{1}{2}\left\langle \sigma _{n,m}+\sigma
_{n,m+1}\right\rangle $ vanishes at $T\rightarrow T_{c}$ with the
critical exponent $1/8$. We note that the phase transition does
not occur if $J_{2}<0$ as well as in the ferromagnetic version of
the model (\ref{H}).

The free energy of the model (\ref{H2d}) is related to the ground
state energy of the Hamiltonian (\ref{H}) by
\begin{equation}
F=4J_{2}E_{0}(h_{x},h_{z})  \label{free_en}
\end{equation}

According to Eqs.(\ref{chi}), (\ref{1d-2d}) and (\ref{free_en})
the zero-field susceptibility in the vicinity of the critical
temperature $T_{c}(0)$ can be obtained using exactly solvable
effective model (\ref{Hmap}). At $\left| T-T_{c}(0)\right| \ll
T_{c}(0)$ the susceptibility is
\begin{equation}
\chi J_{2}=0.1412-0.24\frac{J_{1}}{kT_{c}(0)}\frac{T-T_{c}(0)}{T_{c}(0)}\log
\left( \frac{T-T_{c}(0)}{T_{c}(0)}\right)  \label{chi2d}
\end{equation}

The last equation has a form coinciding with the results obtained
for the two-dimensional antiferromagnetic Ising model by the
series expansion method \cite{sykes}.

\section{Summary}

We have studied the antiferromagnetic Ising chain in the mixed
transverse and longitudinal magnetic field. It was shown that the
quantum phase transition existing in the ITF model remains in the
presence of the uniform longitudinal field. Using the DMRG
simulations we have found the critical line in the ($h_{x},h_{z}$)
plane where the mass gap is closed and the staggered
magnetizations along the $X$ and $Z$ axes vanish. It is found
numerically that the model on the critical line is described by
the conformal field theory with the central charge $c=1/2$, i.e.
it belongs to the universality class of the ITF model.

The scaling behavior in the vicinity of the ITF critical point is
studied in detail. It is shown that the mass gap is proportional
to $h_{z}^{2}$ and the contributions to it are given not only by
the second order correction but also by all other divergent orders
of the perturbation series in $h_{z}$. Besides, the analysis of
the perturbation theory shows that the considered model at
$h_{z}\ll 1$ can be mapped to the effective ITF model with
renormalized parameters depending on $h_{z}$. In a framework of
the effective ITF model the behavior of the susceptibility at
$h_{x}\sim 1/2$ is determined.

The behavior of the model in the vicinity of the multicritical
point is investigated. Using both the variational approach and the
numerical diagonalization results we have found that the mass gap
is proportional to $h_{x}$. Close to the multicritical point the
form of the critical line is linear.

Of course, the considered model is the simplest case of the $XXZ$
model (\ref {HXXZ}). It is interesting to extend the present
analysis to this model as well as to study effects of interchain
interactions in quasi-one-dimensional generalization of the model
(\ref{H}).

The mapping of the quantum model (\ref{H}) to the strongly
anisotropic statistical two-dimensional Ising model in the uniform
magnetic field was considered. The behavior of the susceptibility
of this model near the critical temperature is found. We expect
also, that the main features of the considered model (\ref{H}) are
valid for the statistical two-dimensional antiferromagnetic Ising
model in the uniform magnetic field $h$. That is, the applied
magnetic field does not smear\ the phase transition existing in
the two-dimensional Ising model, which is generic case for the
whole transition line $T_{c}(h)$ in the plane ($T,h$). In
particular, we believe that the analysis of the perturbation
series in $h\ll 1$ in the same manner as was done in Sec.III gives
the scaling behavior for the correlation length and the form of
the critical line near the point $T_{c}(0)$ is
$T_{c}(0)-T_{c}(h)\sim h^{2}$.

This work is supported under RFBR Grant No 03-03-32141 and ISTC No
2207.

\end{document}